\documentclass[
reprint,
superscriptaddress,
showpacs,preprintnumbers,showkeys,
amsmath,amssymb,
aps,
pra,
]{revtex4-1}

\usepackage{graphicx}
\usepackage{dcolumn}
\usepackage{bm}
\usepackage{siunitx}
\usepackage[dvipsnames]{xcolor}

\usepackage[ansinew]{inputenc}
\usepackage{multirow}
\usepackage{stackengine}
\usepackage{capt-of}
\usepackage{bm}
\usepackage{tabularx}

\newcommand{\RM}[1]{\MakeUppercase{\romannumeral #1}}
\newcommand{\varendash}[1][5pt]{\makebox[#1]{\leaders\hbox{--}\hfill\kern0pt}}

\newcommand{\itp}{I. Institute of Theoretical Physics, University of Hamburg, Jungiusstra\ss e 9, D-20355 Hamburg, Germany} 
\newcommand{\ins}{Institute for Nanostructure and Solid State Physics, University of Hamburg, Jungiusstra\ss e 9, D-20355 Hamburg, Germany} 
\newcommand{\dpm}{Department of Theoretical Physics and Applied Mathematics, \\ Ural Federal University, 19 Mira Street, Yekaterinburg, 620002, Russia}

\begin{document}

\title{An atomically thin oxide layer on the elemental superconductor Ta(001) surface}

\author{R. Mozara} \email{rmozara@physnet.uni-hamburg.de} \affiliation{\itp}
\author{A. Kamlapure} \email{akamlapu@physnet.uni-hamburg.de} \affiliation{\ins}
\author{M. Valentyuk} \affiliation{\itp} \affiliation{\dpm}
\author{L. Cornils} \affiliation{\ins}
\author{A. I. Lichtenstein} \affiliation{\itp} \affiliation{\dpm} 
\author{J. Wiebe} \affiliation{\ins}
\author{R. Wiesendanger} \affiliation{\ins}

\date{\today}

\begin{abstract}
Recently the oxygen-reconstructed tantalum surface Ta(001)-p(3$\times$3)-O has experienced considerable attention due its use as a potential platform for Majorana physics in adatom chains. Experimental studies using scanning tunneling microscopy and spectroscopy found rich atomic and electronic structures already for the clean Ta(001)-O surface, which we combine here with \textit{ab initio} methods. We discover two metastable superstructures at the root of the different topographic patterns, discuss its emergence during annealing, and identify the electronic properties. The latter is determined as the sole origin for the contrast reversal seen at positive bias. The observed effects are essentially connected to the two distinct oxygen states appearing on the surface in different geometries. The second superstructure was found in simulations by introducing oxygen vacancies, what was also observed in tantalum pentoxide systems. Additionally we study the charge distribution on the oxidized surface and underline its importance for the adsorption process of polarizable atoms and molecules.
\end{abstract}


\maketitle

The surfaces of elementary superconductors have recently attracted a lot of attention due to their potential in being used as platform for chains which may host Majorana quasiparticles \cite{Nadj-Perge2014, Ruby2015, Ruby2017, Jeon2017, Kim2018, Kamlapure2018}. One requirement for the formation of Majorana states is a strong spin-orbit coupling in the magnetic chain on superconductor system which facilitates the formation of non-collinear magnetization states. Therefore, high-$Z$ elementary superconductors which have an experimentally easily accessible transition temperature above \SI{1}{K} are particularly interesting.

While clean Pb and Re surfaces have been explored \cite{Nadj-Perge2014, Ruby2015, Ruby2017, Jeon2017, Kim2018}, the preparation of clean Ta, La, and Nb surfaces is more challenging \cite{Eelbo2016, Titov1985, Kuznetsov2010, Razinkin2010} particularly due to the tendency to form O reconstructions at the surface. 

On the other hand, such reconstructions also add to the functionality of the surface, as they tend to decouple the spins of adatoms from the substrate conduction electrons \cite{Heinrich2004, Hirjibehedin2007}, which enables to tune the coupling of the adatom spins to the Cooper pairs \cite{Cornils2017}. Ta(110) and Ta(001) have been studied by scanning tunneling microscopy (STM) \cite{Eelbo2016, Cornils2017}. However, the way the structure of the O reconstruction of Ta(100) is linked to the STM images found in Ref.~\cite{Cornils2017} remained elusive.  

A first attempt to characterize the geometry of the Ta(001)-O surface was done by Titov~\textit{et al.} \cite{Titov1985}. Few simple models were proposed to verify the LEED and AES experiment, and it was shown that a couple of O atom arrangements with different coverages can appear at various temperatures. Also, a modified surface with a superimposed (3$\times$3)O network was predicted. Other structures for the oxidized Ta surface were studied recently by Guo~\textit{et al.} \cite{Guo2017} and Bo~\textit{et al.} \cite{Bo2018} from an electronic and quantum-chemical point of view \cite{Sun2014}. In case of Ta(001) \cite{Guo2017}, the study concentrated on O atoms adsorbed at hollow positions only, in contrast to the models proposed by Titov~\textit{et al}. Adsorption at low-coordinated bridge positions, however, was found relevant for bcc metals as regards reconstruction \cite{Koller2001}, catalysis \cite{Brambilla2013}, and CO coadsorption \cite{Zhang2000}. 

\begin{figure}
\includegraphics[width=\columnwidth]{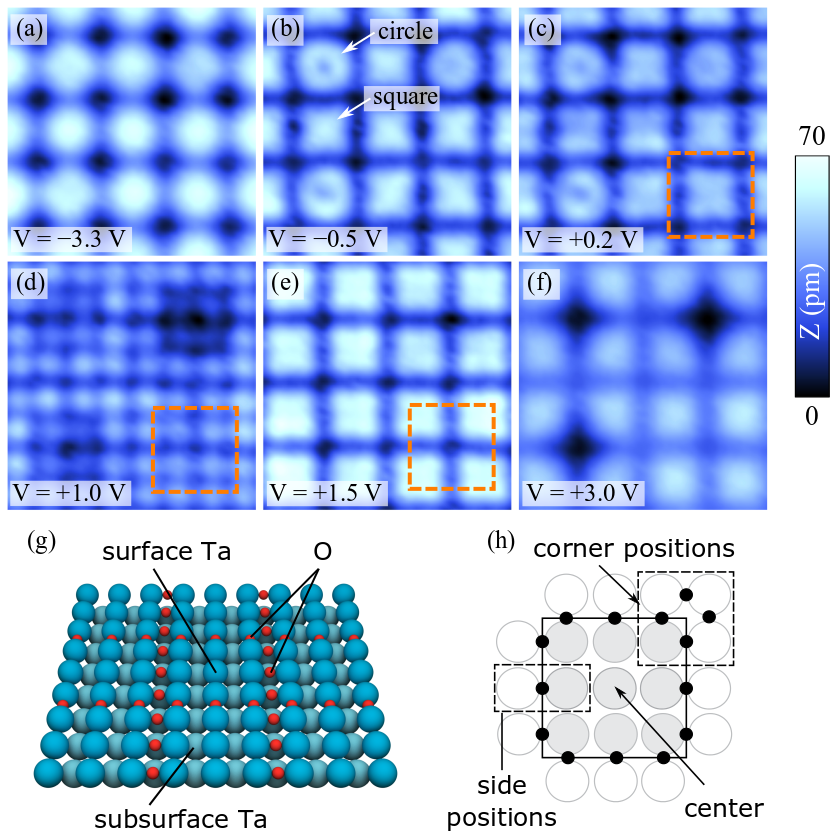}
\caption{\label{fig:topos_exp} (a)-(f) STM images (\SI{3.8}{nm}$\times$\SI{3.8}{nm}) of the O-reconstructed Ta(001) surface measured at the same location, but at various bias voltages as indicated in each panel ($I$\,$=$\,\SI{0.6}{nA}). The dashed square in (c-e) represents the same surface area size of \SI{1.34}{nm}$\times$\SI{1.34}{nm} showing a contrast reversal between $V$\,$=$\,\SI{0.2}{V} and $V$\,$=$\,\SI{1.5}{V}. (g) Schematic perspective view on the surface model indicating the atomic positions as the starting point for the relaxation. (h) Schematic top view indicating the notion of the sites of O and Ta atoms.}
\end{figure}

Here, we present a joint study of Ta(001)-p(3$\times$3)-O by means of experimental (STM, scanning tunneling spectroscopy (STS)) and first-principles (density functional theory (DFT)) techniques. We show the interplay between two types of O positions being the reason for two distinct shapes of 3$\times$3 plaquettes observed in the STM images. Charge transfer between surface sites revealed a distinct polarization texture, which, together with the electronic structure, we predict to be relevant for adsorption of other atoms and molecules.

The surface under study was prepared as described in the Supplementary Information below. To investigate local spectroscopic properties of the sample, $\mathrm dI/\mathrm dV$ spectra were taken using a W tip via Lock-in technique with stabilization voltage and current $V_\mathrm{stab}$ and $I_\mathrm{stab}$, and with a modulation voltage of $V_\mathrm{mod}$ ($f$\,$=$\,\SI{827}{Hz}) added to the sample bias voltage $V$.

Fig.~\ref{fig:topos_exp} shows STM images measured at the same location at various $V$. From Fig.~{\ref{fig:topos_exp}(b)} it can be seen that oxidized Ta(001) forms a well ordered superstructure lattice where a regular network of plaquettes of square and circular shapes separated by continuous depression lines with an apparent depth of $\sim$\,\SI{30}{pm} at $V$\,$=$\,$+$\SI{0.2}{V} are visible. Square-shaped plaquettes are much more frequent than circular-shaped plaquettes with a relative abundance of 4:1. The lateral distance of these plaquettes is \SI{1}{nm} $\sim$~\SI{3}{}$\,a_\mathrm{Ta}$, with the lattice constant $a_\mathrm{Ta}$\,$=$\,\SI{3.3}{\AA} of Ta, which reveals the 3$\times$3 nature of the superstructure formation. The periodicity is consistent with the structure that has been assumed by Titov \textit{et al.} \cite{Titov1985} (see Fig.~\ref{fig:topos_exp}(g)). However, it is \textit{a priori} unknown, whether the plaquettes are due to 3$\times$3 Ta atoms and the depression lines are the O atoms, or vice versa, and what the reason is for the circular- and cross-shaped appearance of the plaquettes. This is further complicated by the second most important experimental result of the present study. There is a shift of the contrast by half of the distance between the plaquettes around a bias voltage of \SI{1}{V} (cf. Figs.~\ref{fig:topos_exp}(c-e)). 

\begin{figure}
\includegraphics[width=\columnwidth]{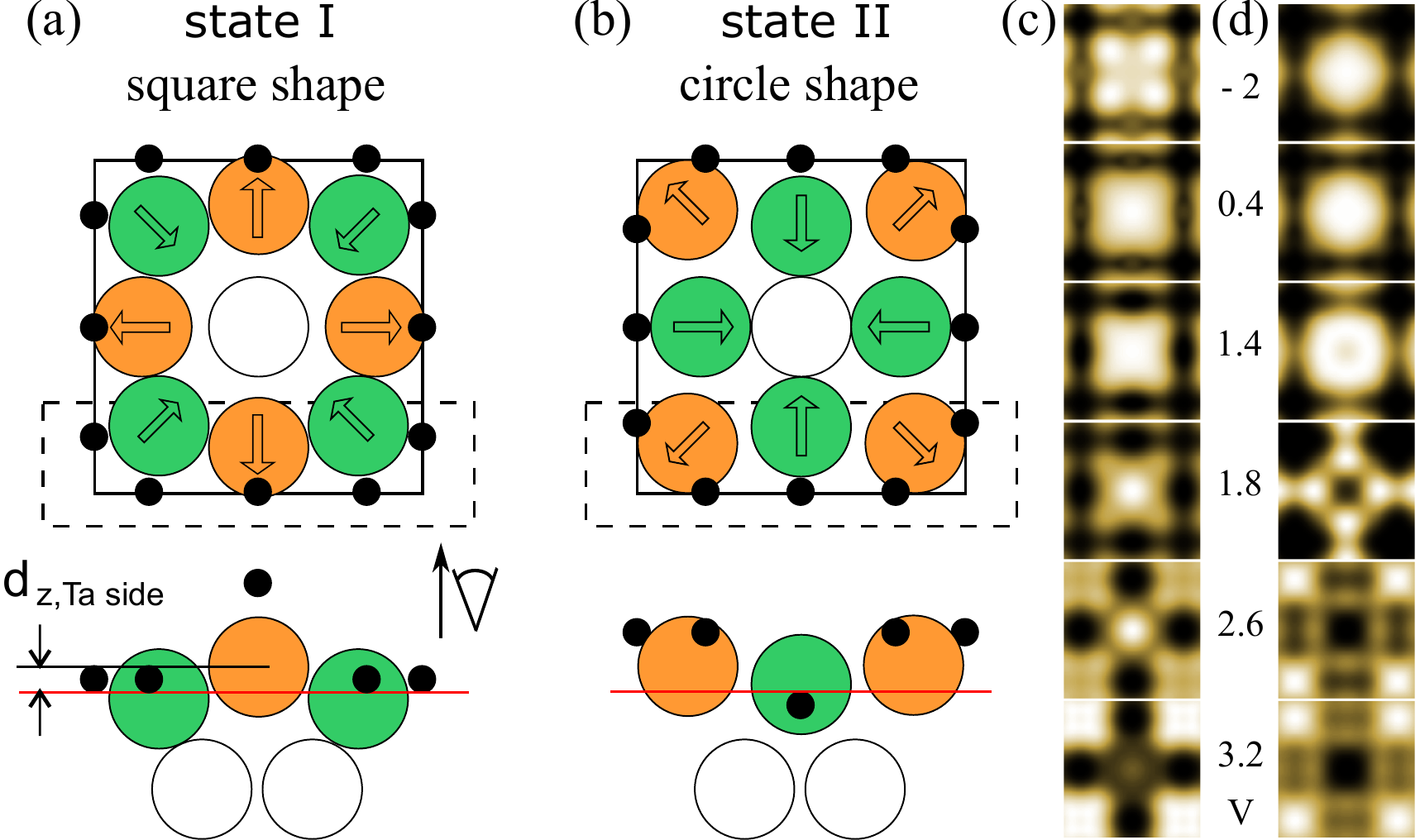}
\caption{\label{fig:struct+topos_theory} (a),(b) Schematic top (top) and side (bottom) views of the two DFT calculated structures~\RM 1 and~\RM 2 indicating the Ta (colored circles) and O (black dots) lateral and vertical positions. The red line depicts the average height of the surface Ta atoms. (c),(d) Simulated STM images for states~\RM 1 and~\RM 2, each at the isosurface value 5$\times$\SI{e-5}{}$\,e$/\AA$^3$. The numbers give the corresponding voltages at which the images have been calculated. The images have been generated by the code given in Ref.~\cite{CSW2013}.}
\end{figure}

To reveal the atomic structure in the experimentally observed 3$\times$3 plaquettes, we performed DFT calculations on the (3$\times$3)O superstructure on Ta(001) with the VASP package \cite{Kresse1996}. For this purpose we used spin-polarized LDA+$U$ and GGA+$U$ functionals with enlarged cut-off energies up to \SI{500}{eV}, including $U$ on O atoms, that has been found relevant in oxide systems \cite{Nekrasov2000} (see Supplementary Information for details). The starting geometry was chosen as proposed by Titov~\textit{et al.} \cite{Titov1985}. It consists of an idealized 3$\times$3 array of O atoms superimposed on the Ta(001) surface. The result of the relaxation is displayed in Fig.~\ref{fig:struct+topos_theory}(a), and denoted as state~\RM 1 (Supplementary Information). The length of the 3$\times$3 plaquette is estimated as $3$\,$a_\mathrm{Ta}$\,=\,\SI{9.9}{\AA}, which is consistent with the periodicity of the reconstruction observed in STM. The first interlayer spacing is compressed by \SI{3.4}{\%} in comparison to bulk values. The surface exhibits both out-of-plane and in-plane reconstruction with buckling of Ta atoms, and zigzag-ordered (along $z$-axis) rows of O atoms.

There exists a second energetic minimum for the considered superstructure, which we uncovered accidentally. To obtain this state one should add additional four O atoms around the central Ta atom (see Fig.~\ref{fig:topos_exp}(h) for the notion of the atomic sites), and then relax. Upon removal of these extra atoms and relaxation, the new state, a structural isomer to state~\RM 1, appears. We denote it as state~\RM 2 (Fig.~\ref{fig:struct+topos_theory}(b)). As we show in Figs.~\ref{fig:struct+topos_theory}(c,\,d), and will explain later in the text, states~\RM 1 and~\RM 2 correspond exactly to the cross and circular types of 3$\times$3 plaquettes, which are observed on the STM images. Structural parameters are in agreement with the ones proposed by Titov~\textit{et al.} from LEED and AES experiments.

The revealed zigzag-ordered positions along the O rows are due to a repulsive interaction between the O atoms. In the ideal positions, two adjacent O atoms at the corner have overlap of their Wigner-Seitz spheres, so they act repulsively, especially in state~\RM 2, where they are elevated. As detailed in the Supplementary Information (Tab.~\ref{tab:structure}), after relaxation, the O atoms at the side positions in state~\RM 1 are higher above the surface by \SI{1.19}{\AA} (Fig.~\ref{fig:struct+topos_theory}(a)), and can be viewed as $sp^3$ hybridized in a tetrahedral surrounding. Two vacuum-oriented hybrid orbitals host approximately two lone pairs, and are very large in extent (Tab.~\ref{tab:charges} in the Supplementary Information). The two neighboring O atoms at the corner positions (state~\RM 1) sidestep into the Ta surface to avoid overlap with O atoms nearby, and form the geometry of the $sp$ hybridization (Supplementary Information). In the rest of the paper we use the name `$sp$' to denote this geometry. In state~\RM 2, the zigzag-ordered heights are reversed.

To identify the contrast seen in the experimental STM images, we explored the Tersoff-Hamann (TH) model in an analogous way as done by Klijn~\textit{et al.} \cite{Klijn2003} (see also Supplementary Information). In the rest of the paper we show results obtained within the LDA, as the electronic structure remains essentially unchanged by use of GGA and vdW functionals, what we have checked explicitly. All simulated STM images were evaluated at the charge density isosurface value \SI{5e-5}{}$\,e$/\AA$^3$, corresponding to a tip-to-surface distance of \SI{5.93}{\AA} at $V$\,$=$\,\SI{3.2}{V} if the tip is above the center of the 3$\times$3 plaquette in state~\RM 1. Independence of the STM contrast on the tip height was checked in experiments and in simulations.

Exploration of the TH approach on both, states~\RM 1 and~\RM 2, leads to the simulated topographs in Figs.~\ref{fig:struct+topos_theory}(c,\,d), and the corresponding differential conductances in Figs.~\ref{fig:didus}(k,\,l) in the Supplementary information. For negative and smaller positive bias up to $\sim$\,\SI{1.4}{eV} we observe depressions along the O rows being maximal at the corner positions, and protrusions above the Ta atoms. The latter match perfectly the square (\RM 1) and circular (\RM 2) shapes seen in STM images around the Fermi energy (Fig.~\ref{fig:topos_exp}(b,\,c) and Fig.~\ref{fig:didus} in Supplementary Information), and we therefore conclude that, in this bias regime, the depression lines correspond to the O rows.

Oxygen is typically seen in low-bias STM images as depression as discussed in a number of papers \cite{Sautet1997, Diebold1996, Woolcot2012, Picone2010, Sun2014}. Similar to TiO$_2$ \cite{Diebold1996, Woolcot2012}, the $s$-$d$ states of the transition metal atom decay much slower into the vacuum above the surface as compared to the O states. This effect overcompensates the stronger exposure of the O atoms to the STM tip expected from their position above the Ta atoms.

For larger positive bias above \SI{1.8}{V}, both, for states~\RM 1 and~\RM 2, the simulated STM images show a contrast reversal (Fig.~\ref{fig:struct+topos_theory}(c,d)). The cross-shaped depression relocates to the center and side Ta atoms. This nicely reproduces the contrast reversal observed experimentally around $V$\,$=$\,\SI{1}{V} (c.f. Figs.~\ref{fig:topos_exp}(c-e)). As we will see below, it originates from a redistribution of the electronic density, i.e., less density appears in the vacuum above these Ta atoms at the corresponding bias voltage.

\begin{figure}
\includegraphics[width=\columnwidth]{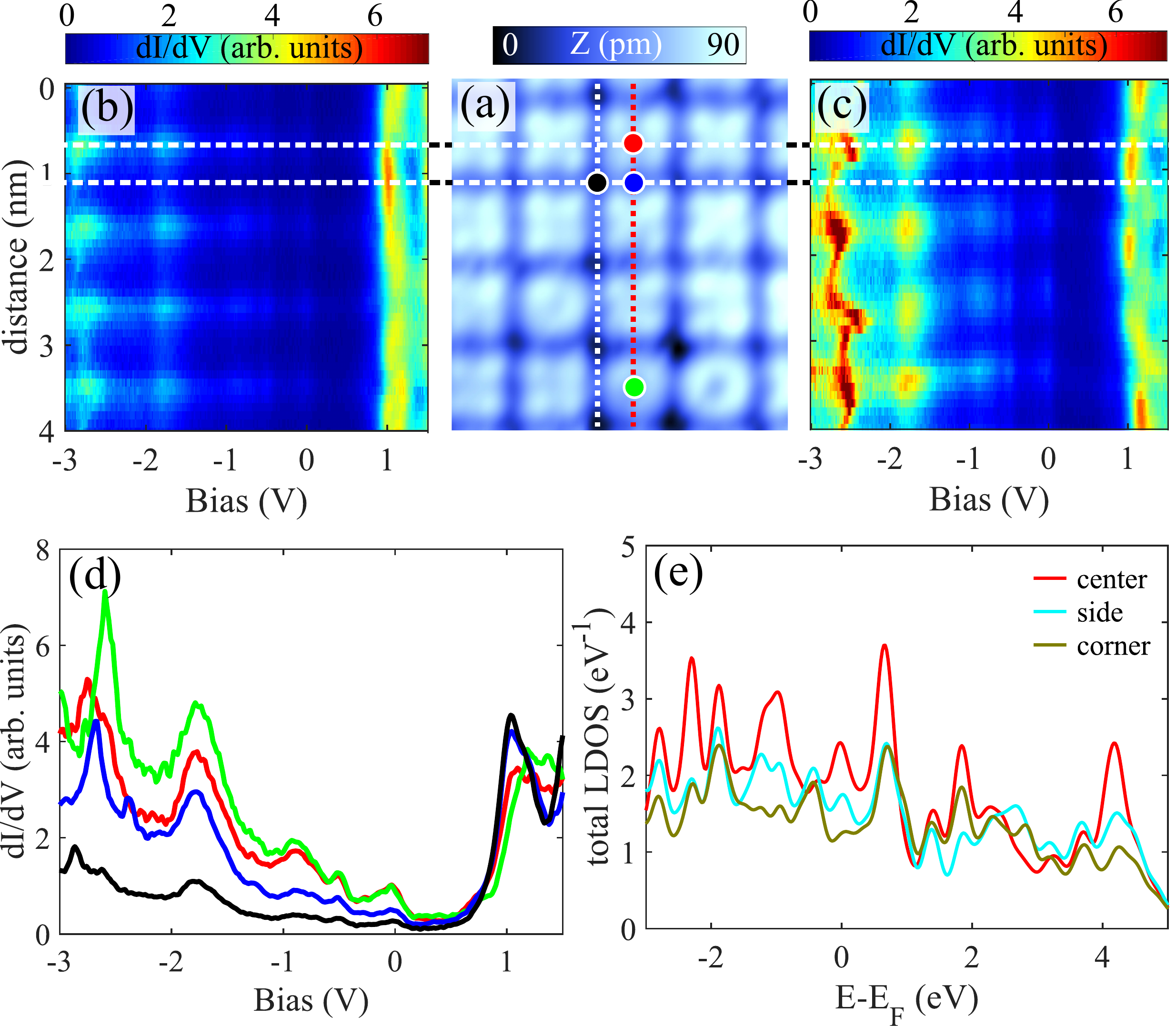}
\caption{\label{fig:spectro_exp+theory} (a) STM image of a surface area used for spectroscopy (\SI{4.0}{nm}$\times$\SI{8.0}{nm}, $V$\,$=$\,\SI{300}{mV}, $I$\,$=$\,\SI{100}{pA}). (b),(c) 2D representation of $\mathrm dI/\mathrm dV$ spectra taken along the white (b) and the red (c) dashed vertical lines marked in (a). (d) Representative $\mathrm dI/\mathrm dV$ spectra acquired at four different locations marked by corresponding colored filled circles in (a). $V_\mathrm{stab}$\,$=$\,\SI{1.5}{V}, $I_\mathrm{stab}$\,$=$\,\SI{0.6}{nA}, $V_\mathrm{mod}$\,$=$\,\SI{5}{mV}. (e) Total DOS of the central, side and corner Ta in state~\RM 1.}
\end{figure}

Fig.~\ref{fig:spectro_exp+theory} shows spatially resolved STS data. Figs.~{\ref{fig:spectro_exp+theory}(b,\,c)} depict 2D colormap representations of STS data acquired along one line on top of the row of O atoms and one line across the Ta plaquettes marked by the white and red dashed lines in Fig.~{\ref{fig:spectro_exp+theory}(a)}, respectively. At negative bias the $\mathrm dI/\mathrm dV$ intensity is largest above the Ta plaquettes ((b,\,c), top horizontal dashed line) and reduced above the O rows (bottom horizontal dashed line). This contrast is reversed at positive sample bias around \SI{1}{eV}, where the $\mathrm dI/\mathrm dV$ intensity is shifted towards the O rows and strongest on the corner O atoms (see bottom horizontal dashed line in (b)). This is further evident from the panel~(d) where we plot spectra taken at the four characteristic locations marked by the correspondingly colored circles in panel~(a). Here, in the negative bias regime, the $\mathrm dI/\mathrm dV$ intensity on the central Ta atoms is larger than that on the side and corner O atoms, while the situation is reversed for positive bias around \SI{1}{V}.

These experimental STS results are compared to the calculated LDOS within empty spheres arranged along the topographic isosurface at a bias of \SI{3.2}{eV} (i.e., above the contrast reversal, see Fig.~\ref{fig:dos_direct_center} in Supplementary Information). The calculated LDOS shows a dominating weight above the Ta as compared to the O atoms up to about \SI{2}{eV}. A shift of the vacuum LDOS from the Ta to the O atoms above $\sim$\SI{2.5}{eV} is in qualitative agreement with the STS results.

In the following, we discuss the appearance of the two metastable O superstructures (\RM 1 and~\RM 2) and the connected features in more detail. The clean Ta(001) surface represents a sparse structure of atoms, and during annealing, the O atoms diffuse to positions providing the strongest bonds. It was shown for Ta(001) that O atoms do not penetrate into the bulk \cite{Drandarov1973}. As shown by our calculations within LDA, during chemisorption an individual O atom would prefer the (probably off-central) pyramidal hollow position as it has lowest energy. The adsorption energy at the two-fold bridge position and in the $sp^3$ state (elevated above the surface) is \SI{0.26}{eV} higher compared to the hollow position, and even \SI{0.50}{eV} higher in the $sp$ state. However, experimentally, after the formation of the reconstruction, we see the O atoms in the bridge positions. This discrepancy is resolved by considering a second O atom nearby. The situation is similar to the CO activation, when in the hollow position the bonding orbitals of an O atom are more saturated, so that the bridge position is preferred as it remains chemically more active, and thus it can react with other O atoms \cite{Zhang2000, Diebold1996}. Indeed, within the calculations we find that two nearby O atoms at bridge positions (along [100] or [010]), both in the $sp^3$ state, have the lowest energy, while one O atom at the bridge ($sp^3$) and one at the closest hollow position is \SI{48}{meV} higher, and both at the hollow positions even \SI{60}{meV} higher in energy. 

The adsorption energy is smaller in state~\RM 1 than in state~\RM 2 by \SI{0.36}{eV} (LDA) (Tab.~\ref{tab:structure} in Supplementary Information). Therefore, the comparatively rare experimental appearance of state~\RM 2 can be explained by the necessity of the specific condition to rearrange bonds inside the plaquette. One of the reasons is an obtained instability of two nearby O atoms along the [110] or [1$\bar 1$0] direction, both in the elevated $sp^3$ state: without any supporting orbital mechanism, one of the two drops immediately into the nearest hollow position. 

\begin{figure}
\includegraphics[width=\columnwidth]{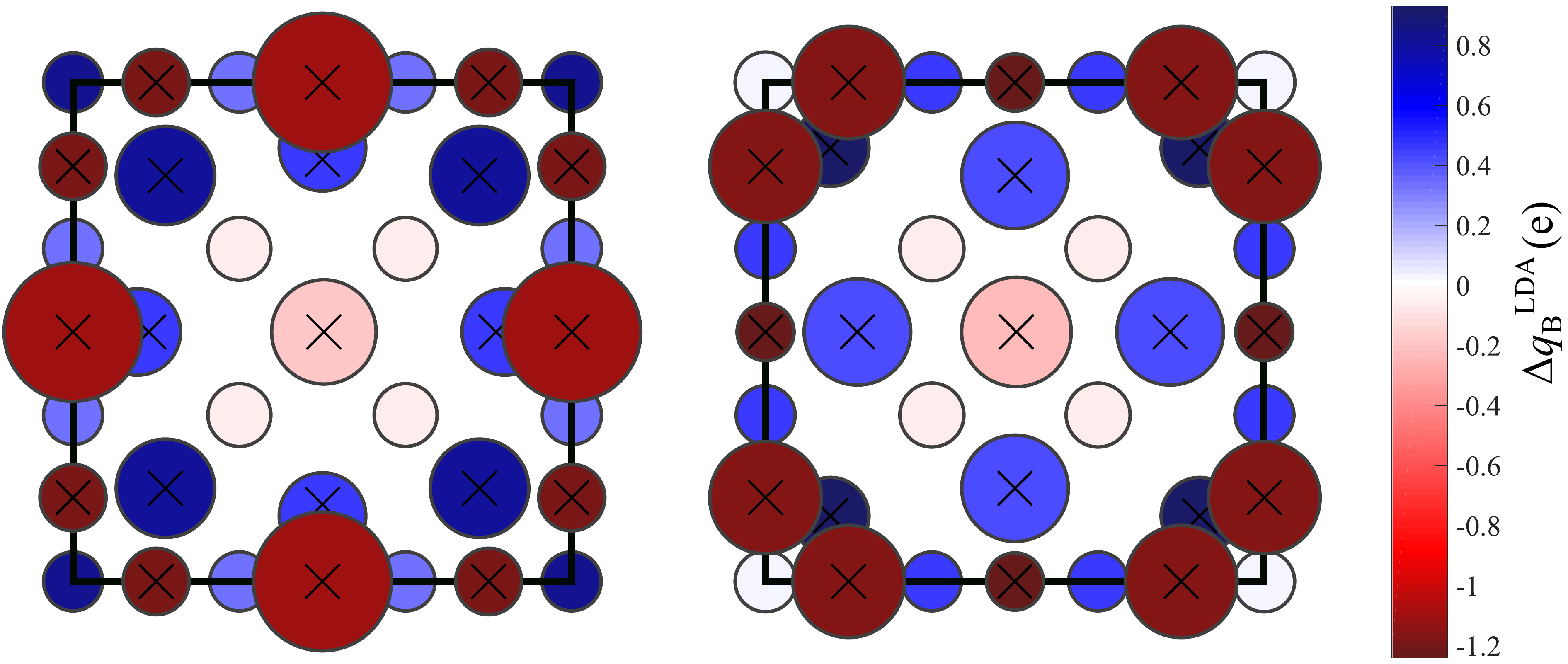}
\caption{\label{fig:charges} Schematic colormap of Bader charges within the LDA for state~\RM 1 (left) and state~\RM 2 (right). Displayed are the differences $\Delta q_\mathrm B$ measured with respect to the free valencies. Blue and red depict charge shortage and excess, respectively. The spheres have been rescaled down by a factor of 2.5 for clearer visibility. Surface atoms are marked by black crosses, to distinguish them from subsurface atoms.}
\end{figure}

We also provide a charge transfer examination within the surface by means of the BCA \cite{Bader} (Fig.~\ref{fig:charges}). A pronounced charge polarization emerges as the O atoms try to reach the O$^{-2}$ state within the $sp$ and $sp^3$ states, which enhances ionicity of all surface atoms.  In general, BCA yields larger estimates for the charge transfer than the iterative Hirshfeld algorithm (which we have checked explicitly with vdW functionals), but still showing similar tendencies. Bu\v{c}ko~\textit{et al.} observed an agreement between the iterative Hirshfeld and Born effective charges for ionic crystals \cite{Bucko2014}. Thus, the occured reconstruction and buckling of the surface Ta atoms are likely to be a polarization-driven distortion.

The BCA result also implies some important considerations for the electronic properties of the surface. We see a pronounced accumulation of negative charge (Fig.~\ref{fig:charges}) at oxygen locations. As a simple approximation, this can be regarded as a local electrostatic potential (ESP) \cite{Kakekhani2018}, additionally acting on the surface near the oxygen rows. Together with filled O states, which are located far below the Fermi level (see Supplementary Information), the ESP will lower the tunneling probability at these sites. Extraordinarily, the central Ta atom is also slightly negatively charged in both structural states. In state~\RM 2 there is less amount of available electronic states at the central Ta atom (Supplementary Information), so it frequently appears as darker spot in STM images. The enhanced ionicity is probably also responsible for the gap opening in the STS curves (Fig.~{\ref{fig:spectro_exp+theory}(d)}).

Large volumes of charge spheres at $sp^3$ locations maps very well to the picture of lone pairs of oxygen (see Supplementary Information for more details). These spheres are very close to the central Ta atom and, taken as lone-pairs, are suggested to induce static dipoles at the nearest atoms \cite{Sun2014}. Indeed, we see enhanced polarized states on the central Ta atom (Fig.~\ref{fig:spectro_exp+theory}(e)). Due to the smearing of the DOS at the tip positions, and the outward and inward curvatures of protrusions and depressions, respectively, the antibonding dipoles can be observed in the STS curves above $E_\mathrm F$ on all regions of the surface, while below $E_\mathrm F$ the peaks are mostly due to the lone pairs \cite{Sun2014} (see also Supplementary Information). The interplay between antibonding electronic states of O and corner Ta atoms with polarized $p$-$d$ states of the central Ta atom is the source of the observed contrast reversal (Supplementary Information).

The pronounced bonding and antibonding peaks in the central Ta states at approximately \SI{-2}{eV}, \SI{-1}{eV}, \SI{0.6}{eV}, \SI{2}{eV} and \SI{4}{eV} (Fig.~\ref{fig:spectro_exp+theory}(e)) can mediate the electrostatic interaction with adsorbates possessing a dipole moment \cite{Kakekhani2018}. This means also, that induced dipole excitations would lead to vdW forces acting between the central Ta atom and such polarizable atoms and molecules. In state~\RM 2, we observe the intensity of polarized peaks being twice lower (Fig.~\ref{fig:dos_direct_center} in Supplementary Information).

In summary, we have unravelled the complex atomic structure of the Ta(001)-p(3$\times$3)-O surface. There is, at least, one extra theoretically uncovered metastable surface superstructure (state~\RM 2). The way we obtained this state points to the presence of oxygen vacancy defects during formation of the surface. The enhanced oxygen vacancy diffusion was shown to be the source of the adaptive crystalline structure in amorphous and crystalline forms of Ta$_2$O$_5$ \cite{Jiang2016}, an experimental prototype for the resistive random access memory. Recently \cite{Lee2017}, also stable polarons were predicted to exist in charged vacancy sites that should affect the carriers mobility. The system under study is presumably a low-dimensional fellow of the tantalum pentoxide family, sharing a common feature: structural diversity driven by vacancies due to the alternation of bent ($sp^3$) and aligned ($sp$) Ta-O-Ta geometries. We also show that calculated static polarization properties of the surface alters in respect to the type of oxygen row in the metastable state. The surface dipoles induced predominantly in state~\RM 1 will have important implications for the adsorption geometry of transition metal adatoms towards the use as a platform for Majorana physics.

\begin{acknowledgments}
We acknowledge financial support from the Deutsche Forschungsgemeinschaft (DFG) through Project No. SFB668-A3 and A1, and the European Union's Horizon 2020 research and innovation program under grant agreement No. 696656\,\,-\,\,GrapheneCore1. Work of RW and AK was funded by ERC via the Advanced Grant ASTONISH (No. 338802). We thank Carmen Herrmann, Elisaveta Kovbasa, Malte Harland, Frank Lechermann, Vladimir Antropov, Sergey Brener and Tom\'{a}\v{s} Bu\v{c}ko for helpful discussions.
\end{acknowledgments}

\onecolumngrid

\begin{center}
\centering
\varendash[300pt] \\[0.5cm]
{\large\textbf{Supplementary Information}}
\end{center}

\section{Surface preparation}

The Ta(001)-p(3$\times$3)-O surface was prepared by first sputter cleaning of a single crystal of Ta(001) using high energy (\SI{2}{keV}) Ar$^+$ ions, followed by repeated cycles of annealing at \SI{1250}{}$^\circ$C in presence of O atmosphere (1$\times$\SI{e-6}{mbar}) and flashing up to \SI{2000}{}$^\circ$C. The sample was then transferred into the STM, which has a base pressure of 5$\times$\SI{e-11}{mbar}, where it was cooled down to the base temperature of $T$\,$=$\,\SI{1.1}{K} \cite{Cornils2017}.

\section{DFT calculations}

DFT was done in the framework of the VASP package with the projector augmented-wave (PAW) basis set \cite{Kresse1996, Bloechl1994}. To avoid mirror polarizations due to the supercell repetition, the atomic arrangement was chosen mirror symmetric to the central layer within a slab of five 9$\times$9 Ta layers. In this way the supercell contains 12 O and 45 Ta atoms. The Brillouin zone was covered by a 6$\times$6$\times$1 $\Gamma$-centred $\mathbf k$-point mesh, and convergence with respect to the number of $\mathbf k$-points was tested. All calculations were performed magnetically, and the surface turned out to be non-magnetic everywhere and in all cases considered. For the exchange-correlation energy two different choices were made with the local-density approximation (LDA) and the generalized gradient approximation (GGA) in the variant of Perdew-Burke-Ernzerhof (PBE) \cite{Perdew1996}. We included local Coulomb correlation via DFT+$U$ from the outset by choosing $U = \SI{6}{eV}$ and $J = \SI{0.8}{eV}$ for O atoms, which was found to be relevant for oxidized transition-metal surfaces \cite{Nekrasov2000}. Relaxations were performed until forces were below 5$\times$\SI{e-3}{eV/\AA}. For the calculation of the DOS the number of bands was increased from around 372 (LDA) or 210 (GGA) to around 400 bands, and the energy cutoff from \SI{400}{eV} to \SI{500}{eV}, to assure an accurate description of the states above the Fermi energy, and which smoothed the simulated STM maps (see Sec.~``Differential conductances and topographs'' below). These were generated with a Mathematica code available online \cite{CSW2013}.

\section{Differential conductances and topographs}

For the simulated STM maps, we applied the Tersoff-Hamann model in an analogous way as done by Klijn \emph{et al.} \cite{Klijn2003}, where at sufficiently low voltages the $\mathrm dI/\mathrm dV$ signal was related to the LDOS of the surface by
\begin{equation}
\label{eq:didu_const}
\frac{\mathrm dI(V,x,y)}{\mathrm dV} \propto e \, \rho_\mathrm t(0) \rho_\mathrm s(eV, x ,y) T(eV,V,z).
\end{equation}
Here, the tip DOS $\rho_\mathrm t$ is assumed to be constant around and between the Fermi levels of tip and substrate, and $T(E,V,z)$\,$=$\,$e^{-2\kappa(E,eV)z}$ is the transmission coefficient, with $\kappa$ the decay rate. The surface is covered by the $x$-$y$ coordinates, $z$ is the perpendicular distance of the tip, and $V$ the applied bias voltage between tip and surface. The constant-current topography $z(x,y)$ is obtained by considering the tunneling current
\begin{equation}
\label{eq:topo_const}
I(x,y) \propto \int_0^{eV} \mathrm dE \, \rho_\mathrm s(E,x,y) T(E,V,z(x,y))
\end{equation}
at a fixed value. The quantity on the r.h.s. is identical to the integrated LDOS of the surface at the position of the tip. Plugging the constant-current topography $z(x,y)$ into Eq.~\eqref{eq:didu_const}, one obtains the $\mathrm dI/\mathrm dV$ signal at constant current.

\section{Oxygen adsorption and tantalum reconstruction}

Tab.~\ref{tab:structure} contains the relevant parameters of the (3$\times$3)O superstructure on Ta(001). The termination of the Ta crystal leads to a compression of the outer surface layers: in the LDA, the first interlayer distance differs from the second interlayer distance by only \SI{0.96}{\%}, but from the bulk value $a_\mathrm{Ta}/2$ by \SI{3.35}{\%}. For the clean Ta surface (not shown), the first to second interlayer distance ratio is \SI{16.03}{\%}, which is already near the ratio between the first interlayer distance and the bulk value $a_\mathrm{Ta}/2$ of \SI{16.72}{\%} (for comparison, a compression of \SI{10}{\%}\,$\pm$\,\SI{3}{\%} was reported in Ref.~\cite{Bartynski1989}; and our GGA calculations yield a first to second interlayer distance ratio of \SI{14.12}{\%} for clean Ta(001) (also not shown)). The O adsorption thus reduces the compression of the Ta surface layers considerably.

The collection of surface atoms has been decomposed into trimers (a collection of three atoms) containing one O and its adjacent two Ta atoms, or one O$_\mathrm{side}$ and two O$_\mathrm{corner}$, or one Ta$_\mathrm{side}$ and two Ta$_\mathrm{corner}$. As can be seen from the surface to subsurface interlayer distances $d_z$, the GGA yields a slightly decompressed surface structure and overall greater bond lengths compared to the LDA. This is a well-known underestimation of the bond-lengths inside L(S)DA \cite{Haas2009}. Accordingly, the adsorption energies in the GGA are smaller than in the LDA, but the differences in $E_\mathrm{ads}$ of the two metastable states are nearly the same for both functionals: \SI{0.36}{eV} in the LDA and \SI{0.35}{eV} in the GGA (likewise for the total energies $E_\mathrm{DFT}$: \SI{4.32}{eV} in the LDA, \SI{4.24}{eV} in the GGA). As the bonds are generally larger in the GGA, the angles in all trimers of the surface layer are sharper.

The strongest variation is the interchange of the O atom heights between state~\RM 1 and~\RM 2 (i.e. $d_z$ of the O$_\mathrm{corner}$-O$_\mathrm{side}$-O$_\mathrm{corner}$ trimer, highlighted in blue). While the adsorption height of O$_\mathrm{side}$ above the surface is large in state~\RM 1, in state~\RM 2 O$_\mathrm{corner}$ is elevated (both in gray). Furthermore, the O$_\mathrm{corner}$-O$_\mathrm{side}$ distance $d$ is considerably decreased in State~\RM 2 (highlighted in green), leading to a potential overlap of their Wigner-Seitz spheres with radius \SI{1.55}{\AA}. The same applies to the distance between two adjacent O$_\mathrm{corner}$ in state~\RM 1. The distance between the O$_\mathrm{corner}$ in the structure with ideal positions as shown on Figs.~{\ref{fig:topos_exp}(g)} and~(h) of the main manuscript is rather small with \SI{2.33}{\AA}. The reconstruction within the Ta surface layer is illustrated by the angles $\theta$ showing substantial deviations from 180$^\circ$, the vertex of which point in different directions depending on the state (Fig.~\ref{fig:struct+topos_theory} of the main manuscript). Furthermore, there appears Ta buckling, as can be seen from the corresponding distances $d_z$, which is also indicated in Figs.~{\ref{fig:struct+topos_theory}(a)} and~(b) of the main manuscript.

\begin{table}[t]
\begin{ruledtabular}
\begin{tabular}{cccccccc}
& &\multicolumn{3}{c}{state~\RM 1 (square-shaped)}&\multicolumn{3}{c}{state~\RM 2 (circle-shaped)} \\[0.05cm]
\cline{3-5} \cline{6-8}
\stackon{$E_\mathrm{DFT}$/$E_\mathrm{ads}$ (eV)}{functional +} & trimer~~~~~ &$d$ (\AA)&$d_z$ (\AA)&$\theta$ ($^\circ$)&$d$ (\AA)&$d_z$ (\AA)&$\theta$ ($^\circ$) \\[0.05cm]
\hline \\[-0.25cm]
\multirow{6}{*}{\shortstack[1]{LDA \\ -617.84/7.43 (\RM 1)\phantom{I} \\ -622.16/7.79 (\RM 2)}}
& \rlap{Ta$_\mathrm{side}$} \phantom{Ta$_\mathrm{surfac}$}\ \llap{-}\ \rlap{O$_\mathrm{side}$} \phantom{Ta$_\mathrm{surfa}$}\ \llap{-}\ \rlap{Ta$_\mathrm{side}$} \phantom{Ta$_\mathrm{surface}$}
& 1.92 & \phantom{-}\colorbox{lightgray}{1.19} (1.25, \phantom{-}0.06) & 103.73 & 1.96 & \colorbox{white}{-0.11} (-0.09, 0.02) & 186.56\\
& \rlap{Ta$_\mathrm{corner}$} \phantom{Ta$_\mathrm{surfac}$}\ \llap{-}\ \rlap{O$_\mathrm{corner}$} \phantom{Ta$_\mathrm{surfa}$}\ \llap{-}\ \rlap{Ta$_\mathrm{corner}$} \phantom{Ta$_\mathrm{surface}$}
& 1.97 & \colorbox{white}{\phantom{-}0.08} (0.03, -0.05) & 171.12 & 1.95 & \phantom{-}\colorbox{lightgray}{0.72} (\phantom{-}0.77, 0.05) & 106.41\\
& \rlap{Ta$_\mathrm{corner}$} \phantom{Ta$_\mathrm{surfac}$}\ \llap{-}\ \rlap{Ta$_\mathrm{side}$} \phantom{Ta$_\mathrm{surfa}$}\ \llap{-}\ \rlap{Ta$_\mathrm{corner}$} \phantom{Ta$_\mathrm{surface}$}
& \colorbox{SpringGreen}{3.02} & \colorbox{white}{\phantom{-}0.11} (0.06, -0.05) & 162.30 & \colorbox{SpringGreen}{3.41} & \colorbox{white}{-0.03} (\phantom{-}0.02, 0.05) & 193.39\\
& \rlap{O$_\mathrm{corner}$} \phantom{Ta$_\mathrm{surfac}$}\ \llap{-}\ \rlap{O$_\mathrm{side}$} \phantom{Ta$_\mathrm{surfa}$}\ \llap{-}\ \rlap{O$_\mathrm{corner}$} \phantom{Ta$_\mathrm{surface}$}
& \colorbox{SpringGreen}{3.35} & \phantom{-}\colorbox{SkyBlue}{1.22} (1.25, \phantom{-}0.03) & 137.28 & \colorbox{SpringGreen}{2.62} & \colorbox{SkyBlue}{-0.86} (-0.09, 0.77) & 218.47\\[0.01cm]
\cline{2-8} \\[-0.30cm]
& \rlap{Ta$_\mathrm{center}$} \phantom{Ta$_\mathrm{surface}$}\ \llap{-}\phantom{-}\ \rlap{Ta$_\mathrm{surface}$} \phantom{Ta$_\mathrm{subsurface}$}
&  & \colorbox{white}{-0.04} \phantom{(0.03, -0.05)} &  &  & \colorbox{white}{-0.26} \phantom{(0.03, -0.05)} & \\
& \rlap{Ta$_\mathrm{surface}$} \phantom{Ta$_\mathrm{surface}$}\ \llap{-}\phantom{-}\ \rlap{Ta$_\mathrm{subsurface}$} \phantom{Ta$_\mathrm{subsurface}$}
&  & \colorbox{white}{\phantom{-}1.60} \phantom{(0.03, -0.05)} &  &  & \colorbox{white}{\phantom{-}1.54} \phantom{(0.03, -0.05)} & \\
\hline \\[-0.25cm]
\multirow{6}{*}{\shortstack[1]{GGA \\ -567.06/6.47 (\RM 1)\phantom{I} \\ -571.30/6.82 (\RM 2)}}
& \rlap{Ta$_\mathrm{side}$} \phantom{Ta$_\mathrm{surfac}$}\ \llap{-}\ \rlap{O$_\mathrm{side}$} \phantom{Ta$_\mathrm{surfa}$}\ \llap{-}\ \rlap{Ta$_\mathrm{side}$} \phantom{Ta$_\mathrm{surface}$}
& \phantom{-}1.95 & \colorbox{lightgray}{1.23} (1.30, \phantom{-}0.07) & 101.61 & 2.00 & \colorbox{white}{-0.15} (-0.13, 0.02) & 188.55 \\
& \rlap{Ta$_\mathrm{corner}$} \phantom{Ta$_\mathrm{surfac}$}\ \llap{-}\ \rlap{O$_\mathrm{corner}$} \phantom{Ta$_\mathrm{surfa}$}\ \llap{-}\ \rlap{Ta$_\mathrm{corner}$} \phantom{Ta$_\mathrm{surface}$}
& \phantom{-}2.00 & \colorbox{white}{0.06} (0.00, -0.06) & 170.46 & 1.97 & \phantom{-}\colorbox{lightgray}{0.83} (\phantom{-}0.89, 0.06) & 104.37 \\
& \rlap{Ta$_\mathrm{corner}$} \phantom{Ta$_\mathrm{surfac}$}\ \llap{-}\ \rlap{Ta$_\mathrm{side}$} \phantom{Ta$_\mathrm{surfa}$}\ \llap{-}\ \rlap{Ta$_\mathrm{corner}$} \phantom{Ta$_\mathrm{surface}$}
& \phantom{-}\colorbox{SpringGreen}{3.00} & \colorbox{white}{0.12} (0.06, -0.06) & 160.78 & \colorbox{SpringGreen}{3.42} & \colorbox{white}{-0.04} (\phantom{-}0.02, 0.06) & 194.73 \\
& \rlap{O$_\mathrm{corner}$} \phantom{Ta$_\mathrm{surfac}$}\ \llap{-}\ \rlap{O$_\mathrm{side}$} \phantom{Ta$_\mathrm{surfa}$}\ \llap{-}\ \rlap{O$_\mathrm{corner}$} \phantom{Ta$_\mathrm{surface}$}
& \phantom{-}\colorbox{SpringGreen}{3.37} & \colorbox{SkyBlue}{1.30} (1.30, \phantom{-}0.00) & 134.74 & \colorbox{SpringGreen}{2.72} & \colorbox{SkyBlue}{-1.02} (-0.13, 0.89) & 224.31\\[0.01cm]
\cline{2-8} \\[-0.30cm]
& \rlap{Ta$_\mathrm{center}$} \phantom{Ta$_\mathrm{surface}$}\ \llap{-}\phantom{-}\ \rlap{Ta$_\mathrm{surface}$} \phantom{Ta$_\mathrm{subsurface}$}
&  & -0.04 \phantom{(0.00, -0.06)} &  &  & -0.30 \phantom{(0.00, -0.06)} & \\
& \rlap{Ta$_\mathrm{surface}$} \phantom{Ta$_\mathrm{surface}$}\ \llap{-}\phantom{-}\ \rlap{Ta$_\mathrm{subsurface}$} \phantom{Ta$_\mathrm{subsurface}$}
&  & \phantom{-}1.67 \phantom{(0.00, -0.06)} &  &  & \phantom{-}1.61 \phantom{(0.00, -0.06)} &
\end{tabular}
\end{ruledtabular}
\caption{\label{tab:structure}
(3$\times$3)O superstructure on Ta(001): Distances and angles between the surface atoms grouped into trimers, together with distances between the central Ta atom and the Ta surface layer, and the first interlayer (average height of all atoms within a layer). Site notion is defined in Fig.~{\ref{fig:topos_exp}(h)} of the main manuscript. Results are listed for each DFT functional together with total energies $E_\mathrm{DFT}$ and adsorption energies $E_\mathrm{ads}$\,$=$\,$(N E_\mathrm O$\,$+$\,$E_\mathrm{\mathrm{Ta(001)}}$\,$-$\,$E_{\mathrm{Ta(001)-O}})/N$ with $N = 12$ the number of O atoms in the supercell. The sign for $d_z$ in state~\RM 1 is the reference for the one in state~\RM 2, and angles above 180$^\circ$ within state~\RM 2 indicate the different direction compared to state~\RM 1, in which their vertex is pointing (cf. Figs.~{\ref{fig:struct+topos_theory}(a)} and~(b) of the main manuscript). Numbers in round brackets denote distances $d_z$ of second and third trimer atoms to Ta$_\mathrm{surface}$.}
\end{table}

\section{Bader charges}

The Bader charge analysis associates charge and atom in a rigid manner, separating the Bader volume by the minima in the charge density around atoms (see Ref.~\cite{Bader} and references therein). Tab.~\ref{tab:charges}, and Fig.~\ref{fig:charges} of the main manuscript, show a pronounced charge transfer from the Ta surface to the O atoms. O atoms acquire more charge as getting closer to the surface, which is known as high electronegativity of oxygen in chemistry. Accordingly, the O atoms do not reach the complete O$^{-2}$ state, as well as the Ta surface atoms do not arrive at the complete Ta$^+$ state either. For both states (I and II) Ta$_\mathrm{corner}$ donate more charge as they are attached to two O atoms. In respect to the functionals, the charge transfer within the system is more pronounced in the LDA than in the GGA, again, due to the increased bonding in the LDA.

More interestingly, while the Bader charge analysis considers the surface region around Ta$_\mathrm{center}$ in both states as negatively charged, the Ta$^\mathrm{sub}_\mathrm{center}$ region is already nearly charge neutral. To some extent, the subsurface Ta layers can be traced back to the almost neutral, but still slightly negative (below $-$\SI{0.1}{}$e$) clean Ta(001) surface and its subsurface, which itself originates from undercoordination due to termination of the crystal. Within the LDA and in state~\RM 1, Ta$^\mathrm{sub}_\mathrm{corner}$ donates a considerable amount of its charge to its four adjacent O$_\mathrm{corner}$ due to the smaller O height in this configuration. In general, we see that the distribution of the charge around Ta$_\mathrm{center}$ is very different between state~\RM 1 and state~\RM 2, that should be of cruicial importance for the adsorption of polarizable atoms and molecules on this surface.

The Bader volumes around each atom were assigned to spheres centered at the atomic sites. The Bader radii show a rather diverse behavior. While the Bader charges are larger in the LDA than in the GGA, the Bader radii are only larger if the O atom is inside the surface. If the O atom is elevated above the surface, the Bader radius becomes smaller in the LDA than in the GGA, although it holds more charge. In contrast to that, for each functional itself Bader charges and radii behave proportionately to each other. The Bader radius is larger for the Ta atoms, of which their adjacent O atom is lying inside the Ta surface. Finally, the Ta surface atoms always have larger Bader radii than the subsurface atoms, because the surface compression is lifted by O adsorption, and surface atoms have access to the vacuum.

In connection with the adsorption of polarizable atoms and molecules, the Bader spheres of O$_\mathrm{side}$ and Ta$_\mathrm{center}$ are nearer to each other in state \RM 1 than in state \RM 2. That contributes to the differences in the Ta$_\mathrm{center}$ DOS between the two states.

\begin{table*}[t]
\begin{ruledtabular}
\begin{tabularx}{0.6\textwidth}{clcccc}
pattern & ion & $\Delta q^\mathrm{LDA}_\mathrm B$ & $\Delta q^\mathrm{GGA}_\mathrm B$ & $R^\mathrm{LDA}_\mathrm B$ & $R^\mathrm{GGA}_\mathrm B$ \\
\hline
\multirow{2}{*}{\shortstack[1]{State~\RM 1 \\ (square)}}
& O$_\mathrm{corner}$               & $-$1.21 & $-$1.14 & 1.74 & 1.66 \\
& O$_\mathrm{side}$                 & $-$1.12 & $-$1.05 & 3.58 & 3.69 \\
& Ta$_\mathrm{corner}$              &   +0.80 &   +0.70 & 2.57 & 2.49 \\
& Ta$_\mathrm{side}$                &   +0.45 &   +0.41 & 2.36 & 2.21 \\
& Ta$_\mathrm{center}$              & $-$0.19 & $-$0.24 & 2.73 & 2.68 \\
\cline{2-6}
& Ta$^\mathrm{sub}_\mathrm{corner}$ &   +0.81 &   +0.75 & 1.57 & 1.59 \\
& Ta$^\mathrm{sub}_\mathrm{side}$   &   +0.32 &   +0.37 & 1.57 & 1.58 \\
& Ta$^\mathrm{sub}_\mathrm{center}$ & $-$0.06 & $-$0.01 & 1.65 & 1.65 \\
\hline
\multirow{2}{*}{\shortstack[1]{State~\RM 2 \\ (circle)}}
& O$_\mathrm{corner}$               & $-$1.18 & $-$1.10 & 2.90 & 2.99 \\
& O$_\mathrm{side}$                 & $-$1.24 & $-$1.17 & 1.50 & 1.49 \\
& Ta$_\mathrm{corner}$              &   +0.91 &   +0.88 & 2.03 & 1.92 \\
& Ta$_\mathrm{side}$                &   +0.41 &   +0.34 & 2.78 & 2.75 \\
& Ta$_\mathrm{center}$              & $-$0.24 & $-$0.26 & 2.86 & 2.73 \\
\cline{2-6}
& Ta$^\mathrm{sub}_\mathrm{corner}$ &   +0.01 &   +0.09 & 1.61 & 1.61 \\
& Ta$^\mathrm{sub}_\mathrm{side}$   &   +0.44 &   +0.37 & 1.56 & 1.59 \\
& Ta$^\mathrm{sub}_\mathrm{center}$ & $-$0.06 & $-$0.03 & 1.68 & 1.69 \\
\end{tabularx}
\end{ruledtabular}
\caption{\label{tab:charges} Bader charges $q_\mathrm B$ ($e$) and radii $R_\mathrm B$ (\AA) obtained within the LDA, where free O has valency $2s^22p^4$, and free Ta $5p^65d^46s^1$; and within the GGA, where free Ta has $5d^46s^1$, and O again $2s^22p^4$. Displayed are the differences $\Delta q_\mathrm B$ measured w.r.t. the free valencies.}
\end{table*}

\section{Differential conductance images}

\begin{figure*}[t]
\includegraphics[width=\textwidth]{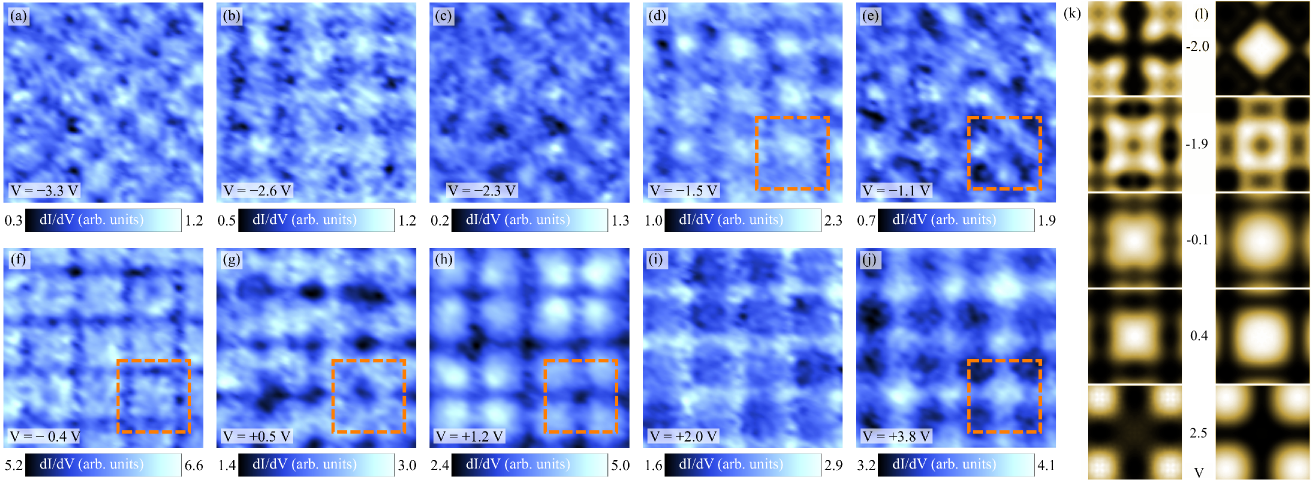}
\caption{\label{fig:didus} (a)-(j) $\mathrm dI/\mathrm dV$ images (\SI{3.8}{nm}$\times$\SI{3.8}{nm}) measured at various bias voltages in the same area as Fig.~\ref{fig:topos_exp}(a)-(f) of the main manuscript ($I_\mathrm{stab}$\,$=$\,\SI{0.6}{nA}, $V_\mathrm{mod}$\,$=$\,\SI{5}{mV}). Dashed squares marked in some of the panels serve as guides to the eyes showing numerous contrast reversals as a function of bias. (k),(l) Simulated $\mathrm dI/\mathrm dV$ images for states~\RM 1~(k) and~\RM 2~(l), each at isosurface value~5$\times$\SI{e-7}{}$e$/\AA$^3$. The integration range around each bias was set to \SI{50}{meV} (same as Gaussian broadening of LDOS in Fig.~\ref{fig:dos_direct_center}). The images have been generated by the code given in Ref.~\cite{CSW2013}.}
\end{figure*}

In Fig.~\ref{fig:didus}, we plot experimental $\mathrm dI/\mathrm dV$ images representative for spatial variations of the DOS at a given bias voltage. Here, the area of measurement for the images is the same as in Fig.~\ref{fig:topos_exp} of the main manuscript. Compared to the STM images, the $\mathrm dI/\mathrm dV$ images show numerous fine variations of the features as a function of bias voltage. In particular, there are at least four contrast reversals similar to the one observed in the STM images, i.e. from $-$\SI{1.5}{V} \ref{fig:didus}(d) to $-$\SI{1.1}{V} (e), from $-$\SI{1.1}{V} (e) to $-$\SI{0.4}{V} (f), from $-$\SI{0.4}{V} (f) to +\SI{0.5}{V} (g) and from +\SI{1.2}{V} (i) to +\SI{3.8}{V} (j). DFT simulation of the conductance leads also to variations of the whole picture, catching contrast reversals in different areas and possible rotation. However, limitations of the TH approach does not allow to discuss it in details. For comparison, we display simulated $\mathrm dI/\mathrm dV$ images of the two states in Fig.~{\ref{fig:didus}(k),(l}). Similar to the experimental $\mathrm dI/\mathrm dV$ images, they reveal multiple contrast reversals, of which we show a selection. However, a one-to-one correspondence between the different features is hampered by the limitations of the TH model.

\section{The Ta$_2$O molecule and orbital hybridization}

The Ta-O-Ta trimers mentioned in Sec.~``Oxygen adsorption and tantalum reconstruction'' comprise (hypothetical) Ta$_2$O molecules. To obtain information on the Ta$_2$O molecule, simplified DFT calculations were performed after singling out the two trimers Ta$_\mathrm{side}$-O$_\mathrm{side}$-Ta$_\mathrm{side}$ and Ta$_\mathrm{corner}$-O$_\mathrm{corner}$-Ta$_\mathrm{corner}$. The DFT setup was left unchanged (see Sec.~``DFT calculations''). Relaxation starting from all trimers in state~\RM 1 and~\RM 2 yields two different molecular geometries (Tab.~\ref{tab:tao_molecule}), which were found to be stable upon perturbations w.r.t. their positions.

The first and second molecule listed in Tab.~\ref{tab:tao_molecule} approximately correspond to the $sp^3$- and $sp$-hybridized configuration, respectively: The bonding angle within the $sp^3$ state is comparable to the one of the H$_2$O molecule (104.45$^\circ$), while in the $sp$ state the trimer with the central O has an almost linear geometry as predicted by the Valence Shell Electron Pair Repulsion (VSEPR) approach \cite{Gillespie1972}. As on the surface, according to the Bader charge analysis the O atoms do not completely reach the O$^{-2}$ state. So the orbitals, which do not participate in bonding, host incomplete lone pairs.

As a molecule itself, the $sp$-hybridized Ta$_2$O has a net magnetic moment of \SI{3.54}{}$\mu_{\mathrm B}$. Furthermore, it is higher in energy by \SI{3.03}{eV} compared to the $sp^3$-hybridized Ta$_2$O molecule. As on the surface, the Bader radius of the O atom in the $sp^3$ state is much larger than in the $sp$ state.

\begin{table}[b]
\caption{\label{tab:tao_molecule}
The Ta$_2$O molecule computed within the LDA: Bond distances $d$ and angles $\theta$, Bader charges $\Delta q_\mathrm B$ and radii $R_\mathrm B$, magnetic moments $\mu$, and total energies $E_\mathrm{DFT}$.}
\begin{ruledtabular}
\begin{tabular}{lccccccccc}
trimer & $d$ (\AA) & $\theta$ ($^\circ$) & $\Delta q_\mathrm B$ ($e$, O) & $\Delta q_\mathrm B$ ($e$, Ta) & $r_\mathrm B$ (\AA, O) & $r_\mathrm B$ (\AA, Ta) & $\mu$ ($\mu_\mathrm B$, O) & $\mu$ ($\mu_\mathrm B$, Ta) & $E_\mathrm{DFT}$ (eV) \\
\hline
Ta\,-\,O\,-\,Ta ($sp^3$) & 1.91 & 111.63 &   $-$0.99 & +0.50 & 2.34 & 2.22 & \phantom{-}0.00 & 0.00 & -19.77 \\
Ta\,-\,O\,-\,Ta ($sp$)   & 1.81 & 179.75 & $-$1.20 & +0.60 & 1.42 & 2.51 & -0.14 & 1.84 & -16.74
\end{tabular}
\end{ruledtabular}
\end{table}

\section{Density of states and surface chemistry}

\begin{figure}
\includegraphics[width=0.6\textwidth]{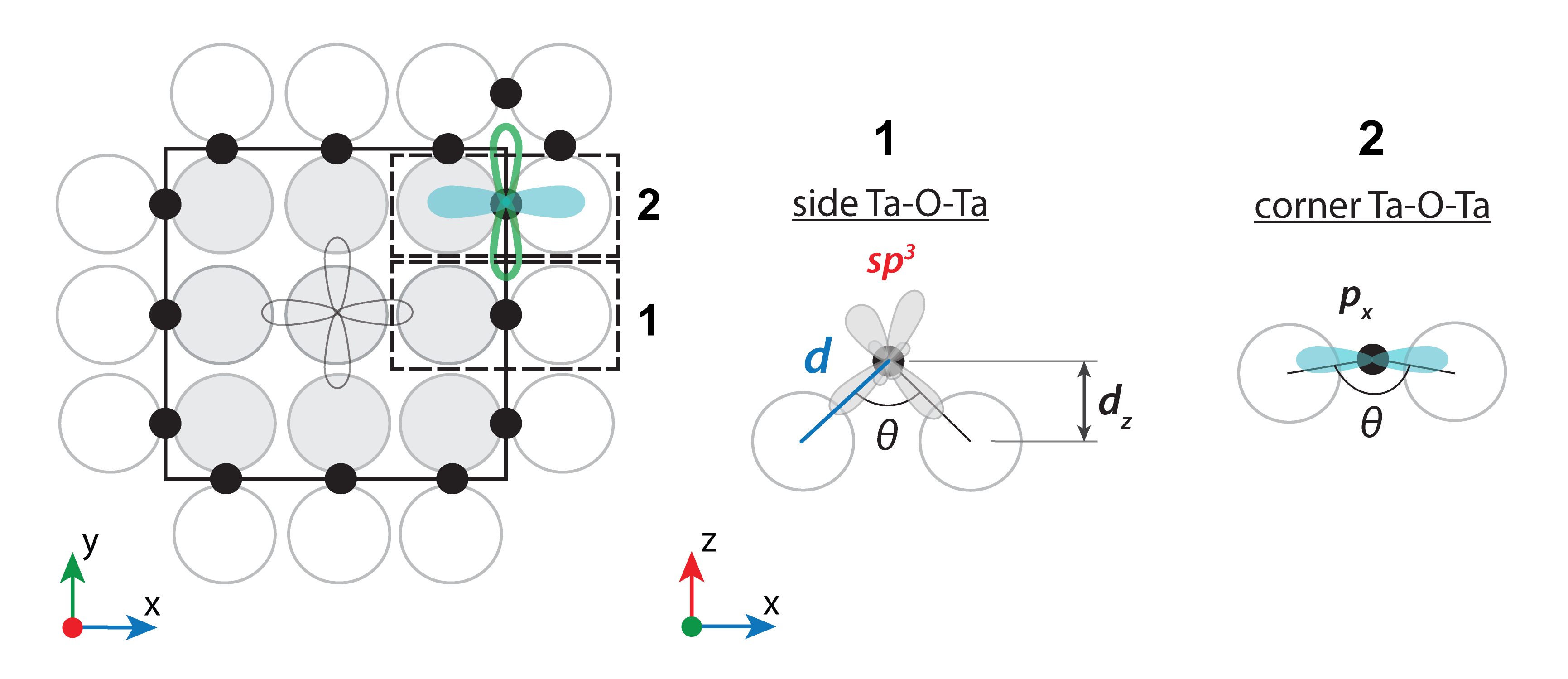}
\caption{\label{fig:struct_orbitals} Idealized surface view with insets showing two configurations of Ta-O-Ta trimers (side views) in state~\RM 1: side group of atoms (1) and corner group (2). An overlay of orbital sketches shows principal difference in electronic states: in light green the ``non-bonding'' $p_y$ orbital of the O$_\mathrm{corner}$ in the vertical row, and in blue its bonding $p_x$ orbital; in (1) we show the proposed $sp^3$ hybridization state for O$_\mathrm{side}$. Color code of orbitals and positions of the atoms under consideration corresponds to Fig.~\ref{fig:dos_direct_1}. For state~\RM 2, O elevations are reversed.}
\vspace{0.75cm}
\includegraphics[width=0.8\textwidth]{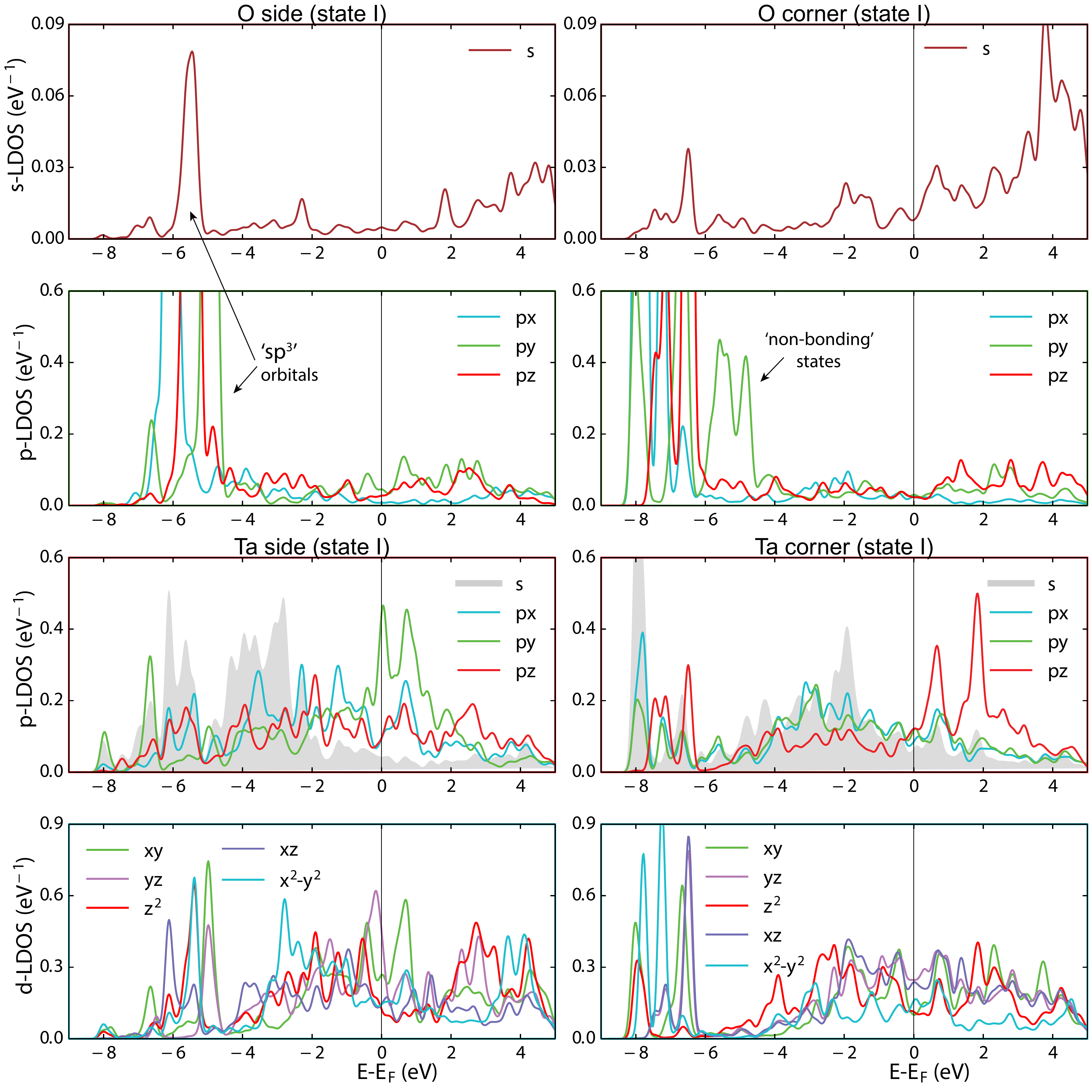}
\caption{\label{fig:dos_direct_1} Orbital-resolved LDOS of Ta(001)-O in state~\RM 1 (LDA). Only the LDOS of atoms sitting on, or next to the vertical rows are shown. See discussion in the text.}
\end{figure}

\textbf{Characterization of the LDOS in state~\RM 1:}
Fig.~\ref{fig:dos_direct_1} contains the orbital-resolved DOS projected onto the sites of the O and corresponding nearest Ta atoms within state~\RM 1. We identify the O $p$ states with energies between \SI{-7.4}{eV}/\SI{-8.3}{eV} and \SI{-5.2}{eV}/\SI{-6.2}{eV} of O$_\mathrm{side/corner}$ as bonding states (Ref.~\cite{Sun2014}), which can be explained by the resonance peaks lying within the same energy range for corresponding peaks in Ta $p$ and $d$ orbitals. 

Let us have a closer look on the DOS of O$_\mathrm{side}$ and O$_\mathrm{corner}$ which are lying on the vertical O row as shown in Fig.~\ref{fig:struct_orbitals}. The LDOS of O$_\mathrm{corner}$ shows a set of non-bonding  states between \SI{-6.2}{eV} and \SI{-4.8}{eV} arising from its $p_y$ orbital (Ref.~\cite{Sun2014} provides a classification of states seen in LDOS of O adsorbates on transition metal surfaces). This can be understood from geometry: The $p_y$ orbital of O$_\mathrm{corner}$ points to hollows. The $p_x$ orbitals of all O atoms on the vertical O rows are completely filled and contribute only to bonding states with adjacent Ta atoms. 

We can apply some quantum chemistry considerations here. According to the Pauling scale of electronegativity, the difference of values between Ta and O in the Ta-O bond reaches 2.0 units, which is close to the border between covalent ($<2$) and ionic ($>2$) type of bonding. So we expect a charge transfer towards O in this bond. Based on the surface trimer geometry, we also identify $sp$ and $sp^3$ hybridized O atoms, as in the hypothetical Ta$_2$O molecules. One should note, that for the $sp$ hybridized O atom embedded into the Ta surface, this localized bonding picture is not relevant, but we use the name, as it carries typical chemical properties. 
Focusing on O$_\mathrm{side}$ in state~\RM 1, one observes $s$ and $p$ states located approximately at the same energy that constitutes the $sp^3$ hybridization. Then, taking into account the saturation of its hybrid orbitals, two of them are directed towards the adjacent Ta$_\mathrm{side}$, and the two remaining point towards the vacuum, along the O row and perpendicular to the other two hybrid orbitals. These host the lone pairs (or non-bonding states), and we argue that these states have important implications for all chemical properties of the system. In the DOS one can see that orbitals of $p_y$ and $p_z$ character are higher in energy, in agreement with the local arrangement. We also identify non-bonding states between \SI{-5}{eV} and up to the Fermi energy, and anti-bonding states above $E_\mathrm F$.

Another type of O atom in state~\RM 1, the $sp$-hybridized O$_\mathrm{corner}$, has a more complex electronic structure. First, the $sp$ state is characterized by a higher electronegativity of the trimer-central atom than the $sp^3$ state, because it has a larger $s$-state contribution (see ``Bader charge section'' in Tab.~\ref{tab:charges}). Second, there is the already mentioned non-bonding set of states for the $p_y$ orbital. It is possible, that this orbital is rotated towards the $sp^3$ orbital of the nearest O$_\mathrm{side}$, being aligned inside the O row. Third, there are clear $\pi$-bonding and anti-bonding features on spectra of neighboring Ta atoms.

\begin{figure}
\includegraphics[height=0.8\textwidth]{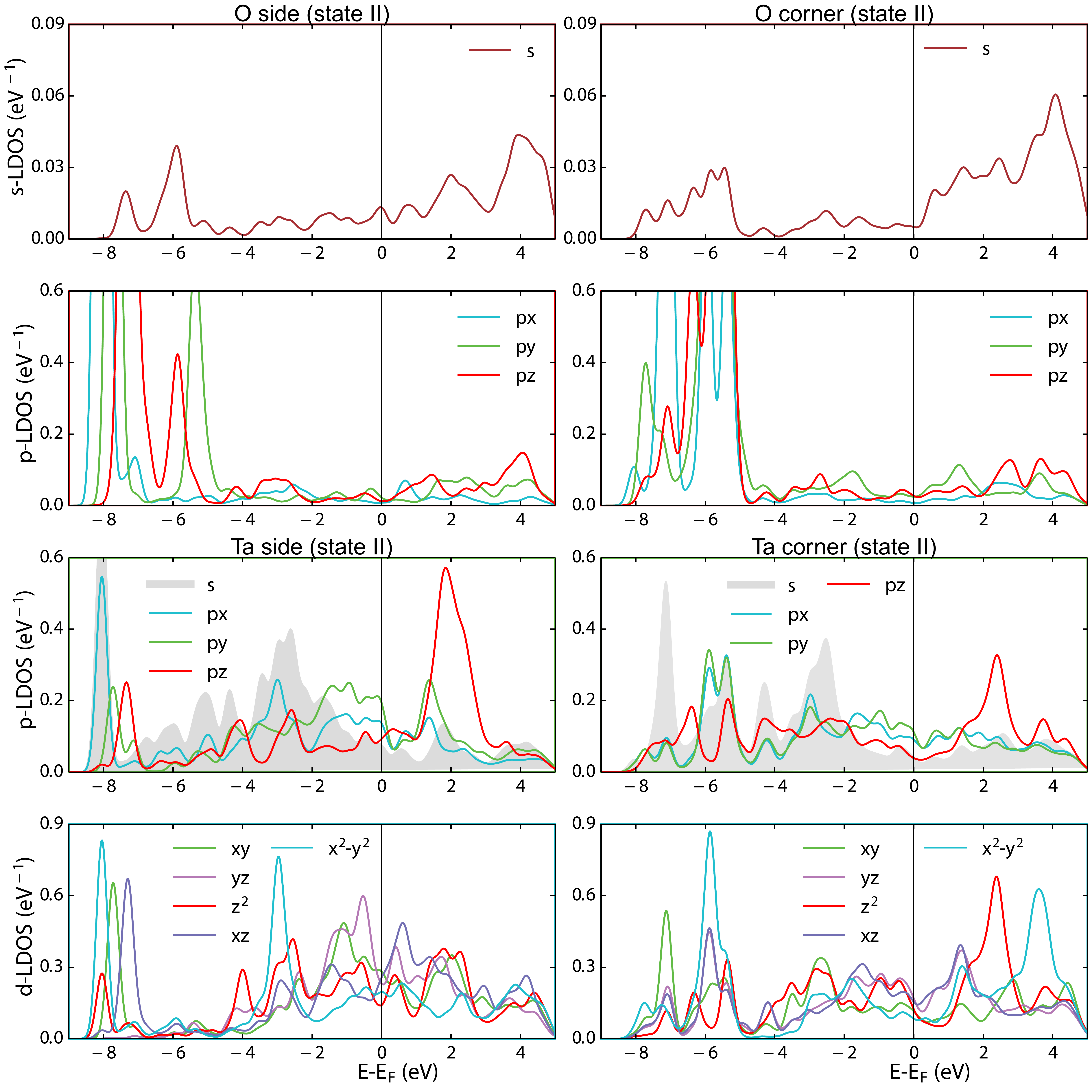}
\caption{\label{fig:dos_direct_2} Orbital-resolved LDOS of Ta(001)-O in state~\RM 2 (LDA). Only the LDOS of atoms sitting on, or next to the vertical rows are shown. See discussion in the text.}
\end{figure}

\vspace{0.3cm}
\textbf{Characterization of the LDOS in state~\RM 2:}
The O-LDOS in Fig.~\ref{fig:dos_direct_2} show a reversed tendency in their behavior, which corresponds to the reversed adsorption heights within the O rows. We suggest that the O$_\mathrm{side}$ $p_y$ and $p_z$ orbitals form a linear combination as $p_y \pm p_z$, with the one lobe pointing to the close-by $sp^3$ orbitals, and the one pointing into the bulk containing the bonding states. Ta$_\mathrm{corner}$ now shows excitations being more pronounced in the $d_{z^2}$ and $d_{x^2-y^2}$ orbitals as compared to Ta$_\mathrm{side}$ in state~\RM 1 (also due to polarizations induced by lone pairs; to be discussed below), because the two O$_\mathrm{corner}$ are now elevated above the surface, and in the analogous state of O$_\mathrm{side}$ within state~\RM 1. The Ta$_\mathrm{side}$ and Ta$_\mathrm{corner}$ $p_z$ orbitals in state~\RM 2 show a reversal in excitation weights compared to state~\RM 1 as well.

\vspace{0.3cm}
\textbf{Effects of $\bm{sp}$ and $\bm{sp^3}$ hybridization on adsorption geometry:}
One can understand the adsorption geometry with the help of the proposed  $sp$ and $sp^3$ hybridizations, and the superstructure provided in Tab.~\ref{tab:structure}, which is depicted in Figs.~{\ref{fig:struct+topos_theory}(a)} and~(b) of the main manuscript. As we mentioned, the two bonding orbitals of the $sp^3$ hybridized O$_\mathrm{side}$ in state~\RM 1 point to the adjacent Ta$_\mathrm{side}$, with both DFT functionals having a slightly smaller bonding angle than the H$_2$O molecule (104.45$^\circ$; Tab.~\ref{tab:structure}). The two remaining lone pairs point diagonally with their orbital lobes to the vacuum along the O row to form the tetrahedral structure.

According to the Bader charge analysis (Sec.~``Bader charges'' above), the orbital volumes holding the lone pairs are very large, and interact repulsively with O$_\mathrm{corner}$, which is then embedded into the Ta surface and forms the geometry of the $sp$ hybridization (the in-plane $p_x$ orbital hybridizes with atoms nearby). The lobe of its non-bonded $p_y$ orbital parallel to the Ta surface occupies the free space along the (vertical) O row and below the lobes of the $sp^3$-hybridized orbitals of O$_\mathrm{side}$ pointing to the vacuum. The geometric situation reverses for State~\RM 2. As there are more $sp^3$-hybridized O atoms than $sp$-hybridized ones in State~\RM 2, it has a lower total energy compared to State~\RM 1 (Tab.~\ref{tab:structure}). This conclusion can also be drawn from Tab.~\ref{tab:tao_molecule}, which shows that the Ta$_2$O molecule in $sp^3$-hybridized configuration has lower energy. To sum up, the repulsive character of the formed oxygen lone pairs in the $sp^3$ state should play an important role during oxygen adsorption and is responsible for the observed surface pattern with zigzag-ordered O structure perpendicular to the surface. Depending on the temperature and pressure, the O coverage and repulsive interactions are balanced, and crossing O rows are admitted until the optimal 3$\times$3 superstructure emerges.

\begin{figure}
\includegraphics[height=0.5\textheight]{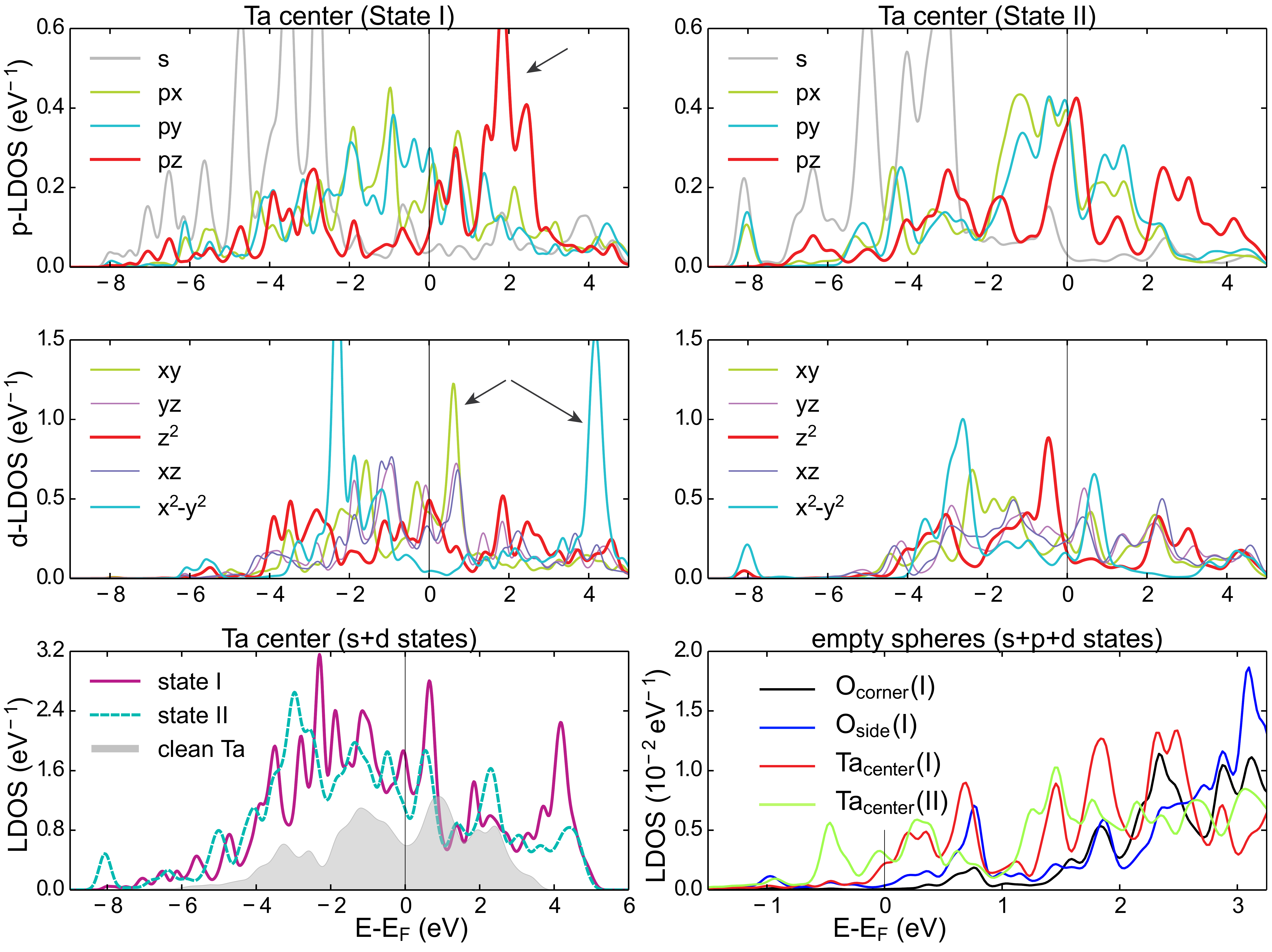}
\caption{\label{fig:dos_direct_center} Orbital-resolved Ta$_\mathrm{center}$-LDOS of Ta(001)-O in states~\RM 1 (top and middle left) and~\RM 2 (top and middle right). Results obtained within the LDA. For state~\RM 1, arrows indicate anti-bonding $\pi$-bond states ($p$-LDOS) and anti-bonding dipole states ($d$-LDOS). The comparison of the overall $s+d$ electronic states of the Ta$_\mathrm{center}$ to the clean Ta surface atom is shown in the bottom left spectrum. LDOS projected onto the added $s$, $p$, and $d$ orbitals of empty spheres above indicated atomic sites in Fig.~\ref{fig:spectro_exp+theory}(a) of the main manuscript, and arranged along the isosurface at \SI{3.2}{eV} shown in Fig.~\ref{fig:struct+topos_theory}(c) of the main manuscript. The radius of the empty spheres has been chosen as the Wigner-Seitz radius of W. All spectra have been broadened using a Gaussian filter with a FWHM of \SI{50}{meV}.}
\end{figure}

\vspace{0.3cm}
\textbf{Induced polarization effects:}
In state~\RM 1, O$_\mathrm{side}$ has approximately two lone pairs in $sp^3$-hybridized orbitals with $p_y$ and $p_z$ contributions. Consequently these can induce dipoles in the Ta$_\mathrm{side}$ orbitals of the same symmetry (the anti-bonding peaks of $p$ states above $E_\mathrm F$), although more pronounced in $p_y$, which is parallel to the two hybrid orbitals carrying the suggested lone pairs, and being aligned along the O row. The effect of polarization by induced dipoles is more pronounced on the Ta surface and subsurface atoms (latter not shown) in the central region of the 3$\times$3 plaquette and below the O row (not shown; for explanations of lone-pair bonding see Ref.~\cite{Sun2014}). Excessive peaks above $E_\mathrm F$ are seen in the Ta$_\mathrm{center}$ $p_z$, $d_{xy}$, $d_{x^2-y^2}$ orbitals (Fig.~\ref{fig:dos_direct_center}), the latter two pointing to O$_\mathrm{side}$ and O$_\mathrm{corner}$. The $p_z$ peaks are due to $\pi$ bonding with the $p_z$ orbital of Ta$_\mathrm{corner}$, that is consistent with local geometry. Pronounced $d_{xy}$ and $d_{x^2-y^2}$ peaks reflect a non-direct hybridization effect that can be cast to the lone-pair idea: these are the lone-pair induced bonding and anti-bonding dipoles. Indeed, the $sp^3$-hybridized states interact with Ta$_\mathrm{side}$ $s$- and $d(p)$-states \cite{Walsh2011}, which in turn induce excessive peaks at $xy$ and $x^2$$-$$y^2$ states of Ta$_\mathrm{center}$.

In state~\RM 2, now the Ta$_\mathrm{side}$ and Ta$_\mathrm{corner}$ $p_z$ orbitals participate in $\pi$ bonding (Fig.~\ref{fig:dos_direct_2}). Ta$_\mathrm{center}$, however, does not show any sharp peaks around and above $E_\mathrm F$, and has significantly less intensity of the whole spectrum in comparison to state~\RM 1 (Fig.~\ref{fig:dos_direct_center}). This can be also attributed to the immersed position of the Ta$_\mathrm{center}$ in state~\RM 2. While the long-range effects of the O atoms on the electron structure is less pronounced, the Ta atoms adjacent to the O row show excessive features above and below $E_\mathrm F$. Especially in the Ta$_\mathrm{corner}$ $d_{x^2-y^2}$ orbital we observe bonding and anti-bonding peaks, which we suggest again to reflect dipoles induced by the presence of two adjacent $sp^3$-hybridized O$_\mathrm{corner}$ atoms.

\vspace{0.3cm}
\textbf{Explanation of plaquette shapes and contrast reversal:}
The alternation of the $sp^3$-hybridized O atoms in corner and side positions along the rows leads to the additional broader darkening at its locations on the maps. This effect tunes the square-shaped ($sp^3$ atom at side positions) and circle-shaped ($sp^3$ atom at corner positions) protrusions  and goes well in line with the lone-pair concept. Also, in state~\RM 2, the $d$-orbital weight of Ta$_\mathrm{center}$ at low bias is rather small, and thus the central dark spot in low-bias STM images appears. This is because the O atom Bader spheres are farer away (cf. Fig.~\ref{fig:charges} in the main manuscript) and lone-pair induced anti-bonding dipole production is low.

Backwards, we define the following rules to identify plaquette shapes. All plaquettes have similar features at small bias: deep minima are at the O$_\mathrm{corner}$, plaquette pattern is present in the system everywhere. We distinguish cross shapes (state~\RM 1) from circle shapes (state~\RM 2) by additional minima/maxima:
\begin{enumerate}
\itemsep0em
\item The $sp^3$-state O site (bent geometry) always shows a minimum (less tunneling probability) and it is delocalized (the minimum spot/area appears broad).
\item The $sp$-state O site (aligned geometry) also always shows a minimum, but a smaller spot.
\item The Ta regions correspond normally to the protrusions.
\end{enumerate}
Then plaquette attributes goes for the cross shape as:
\begin{itemize}
\itemsep0em
\item Additional broad minima are at O$_\mathrm{side}$ sites ($sp^3$),
\item maxima are at Ta$_\mathrm{corner}$ and Ta$_\mathrm{center}$ (not strictly).
\end{itemize}
And for the circle shape as:
\begin{itemize}
\itemsep0em
\item Additional broad minima are at O$_\mathrm{corner}$ sites,
\item maxima are at Ta$_\mathrm{side}$, 
\item and an additional minimum is at Ta$_\mathrm{center}$ (central spot).
\end{itemize}

The contrast reversal means that there are more states above the corner positions at higher biases. There are four Ta$_\mathrm{corner}$ in contrast to only one Ta$_\mathrm{center}$. The antibonding states of O$_\mathrm{corner}$ contribute additionally. For the measurement with the STM tip are the Ta $p_z$ and $d_{z^2}$ orbitals slightly more relevant, and the charge density from different sites tends to interpenetrate at higher distances above the surfaces, especially as anti-bonding states happen to be more delocalized \cite{Jensen1990}. Thus, before the contrast reversal, the STM tip measures the orbitals containing anti-bonding dipole states lowering the work function around Ta$_\mathrm{center}$. At higher bias the anti-bonding states within orbitals of Ta$_\mathrm{corner}$ contribute more efficiently \cite{Sun2014}, also because the potential O-induced gap due to electron-hole production does not prevail anymore.

\vspace{0.3cm}
\textbf{Summary:}
In conclusion, we performed a detailed investigation of the experimentally observed electronic properties of the Ta(001)-O surface by means of DFT. Thus, we were able to identify its contrast reversal as seen in the STM images and to predict the adsorption specific properties. One should mention that DFT-GGA in general has a limited access to electrostatic effects on a surface and overall plays towards its metallic character  (see, for instance, the experimentally observed induced gap (Fig.~\ref{fig:spectro_exp+theory})). The parental bulk material, especially for the local structure containing the $sp^3$-hybridized trimer, is tantalum pentoxide Ta$_2$O$_5$ \cite{Guo2014, Jiang2016}. It is an oxide with a high dielectric constant, and with a reported band gap of $\sim$\,1\,-\,\SI{4}{eV}. DFT-GW simulations improved the gap values for the crystalline form of Ta$_2$O$_5$, although insufficiently \cite{Lee2014}. The contribution of Ta $s$ states in the conductance area for intraplanar bonds was suggested to be the reason for the small gap in $\beta$-Ta$_2$O$_5$ and $\delta$-Ta$_2$O$_5$ \cite{Lee2014}.
  
Also, we found the picture of the localized molecular bonds helpful, namely the one in terms of hybrid orbitals hosting non-bonding electrons (lone pairs), that served as a complementary view. Unexpectedly, we could explain some properties of the surface with repulsive and long-range character of the lone-pairs interaction that coincide with DFT results. 

The Ta protrusions seen on STM images are due to the dominance of the electronic properties over the structural ones. In particular, we propose lone-pair induced anti-bonding dipole $d$ states, and anti-bonding $\pi$-bonded $p$ states should play an important role in the physics of the surface. These states above $E_\mathrm F$ lower the work function, and thus enhance the tunnel current. At higher bias these lead to the contrast reversal.

We see that polarization properties differ in state~\RM 1 and~\RM 2 of the O-reconstructed Ta(001) surface. The arrangement of $sp$- and $sp^3$-hybridized O atoms along the rows induces excitations in the center of the 3$\times$3 plaquette of state~\RM 1 by means of indirect hybridization, but much less in state~\RM 2.  It is O$_\mathrm{side}$ in state~\RM 1, which has the largest Bader radius, and its Bader sphere has the highest proximity to the one of Ta$_\mathrm{center}$. The inhomogeneous electrostatic texture (cf. Fig.~\ref{fig:charges} of the main manuscript) was found to be marked with an unusual formally negative charge on Ta$_\mathrm{center}$. The pronounced bonding and anti-bonding peaks in the surface Ta $d$ states at approximately \SI{-1}{eV}, \SI{-2}{eV}, \SI{0.6}{eV}, \SI{2}{eV} and \SI{4}{eV} can mediate the electrostatic interaction with adsorbates possessing a dipole moment \cite{Kakekhani2018}. This also means that the induced dipole excitations would lead to vdW forces acting between Ta$_\mathrm{center}$ and such polarizable atoms and molecules.  

\twocolumngrid

\bibliography{References}

\end{document}